\documentclass[manuscript]{aastex}

\slugcomment{To appear in ApJ}

\shorttitle{Cometary Activity in (3552) Don Quixote}
\shortauthors{Mommert et al.}

\begin{document}

\title{The Discovery of Cometary Activity in \\ Near--Earth Asteroid (3552)
Don Quixote}

\author{Michael Mommert}
\affil{Institute of Planetary Research, German Aerospace Center (DLR),
  Rutherfordstr. 2, 12489 Berlin, Germany} 
\affil{ Department of Physics and Astronomy, Northern Arizona
  University, PO Box 6010, Flagstaff, AZ 86011, USA}
\author{Joseph L. Hora}
\affil{Harvard-Smithsonian Center for Astrophysics, 60 Garden Street,
  Cambridge, MA 02138-1516, USA} 

\author{Alan W. Harris}
\affil{Institute of Planetary Research, German Aerospace Center (DLR),
  Rutherfordstr. 2, 12489 Berlin, Germany} 

\author{William T. Reach}
\affil{Universities Space Research Association, Stratospheric
  Observatory for Infrared Astronomy, MS 232-11, NASA Ames Research
  Center, Moffett Field, CA 94035, USA} 

\author{Joshua P. Emery} \affil{Department of Earth and Planetary
  Sciences, University of Tennessee, 1412 Circle Dr., Knoxville, TN
  37996, USA}

\author{Cristina A. Thomas}
\affil{NASA Postdoctoral Program Fellow, NASA Goddard Space Flight
  Center, 8800 Greenbelt Rd, Greenbelt, MD 20771, USA}

\author{Michael Mueller}
\affil{SRON Netherlands Institute for Space Research, Postbus 800,
  9700 AV Groningen, The Netherlands} 

\author{Dale P. Cruikshank} 
\affil{NASA Ames Research Center, Moffett Field, CA 94035, USA}

\author{David E. Trilling}
\affil{Department of Physics and Astronomy, Northern Arizona
  University, PO Box 6010, Flagstaff, AZ 86011, USA} 

\author{Marco Delbo'}
\affil{UNS-CNRS-Observatoire de la Cote d'Azur, BP4229 06304 Nice
  Cedex 4, France} 

\and

\author{Howard A. Smith}
\affil{Harvard-Smithsonian Center for Astrophysics, 60 Garden Street, Cambridge, MA 02138-1516, USA}


\begin{abstract}
  The near--Earth object (NEO) population, which mainly consists
    of fragments from collisions between asteroids in the main
    asteroid belt, is thought to include contributions from
    short--period comets as well. One of the most promising NEO
    candidates for a cometary origin is near--Earth asteroid (3552)
    Don Quixote, which has never been reported to show activity.\\
    Here we present the discovery of cometary activity in Don Quixote
    based on thermal--infrared observations made with the Spitzer
    Space Telescope in its 3.6 and 4.5~\micron\ bands. Our
    observations clearly show the presence of a coma and a tail in the
    4.5~\micron\, but not in the 3.6~\micron\ band, which is
    consistent with molecular band emission from CO$_2$. Thermal
    modeling of the combined photometric data on Don Quixote reveals a
    diameter of 18.4$_{-0.4}^{+0.3}$~km and an albedo of
    $0.03^{+0.02}_{-0.01}$, which confirms Don Quixote to be the
    third--largest known NEO.  We derive an upper limit on the dust
    production rate of 1.9~kg~s$^{-1}$ and derive a CO$_2$ gas
    production rate of $(1.1\pm0.1)\cdot 10^{26}$ molecules
    s$^{-1}$. Spitzer IRS spectroscopic observations indicate the
    presence of fine--grained silicates, perhaps pyroxene rich, on the
    surface of Don Quixote. Our discovery suggests that CO$_2$ can be
    present in near--Earth space over a long time. The presence of
    CO$_2$ might also explain that Don Quixote's cometary nature
    remained hidden for nearly three decades. 
\end{abstract}


\section{Introduction}
\label{lbl:intro}

The near--Earth object (NEO) population comprises asteroids and comets
with perihelion distances $q\leq1.3$~AU. As of June 2013, ${\sim}$160
comets and more than 10000 asteroids are known in near--Earth
space\footnote{according to the JPL NEO program:
  \url{http://neo.jpl.nasa.gov/stats/}}. The NEO population is
replenished from collisional fragments from main belt asteroids and
short--period comets \citep[see, e.g.,][]{Wetherill1979, Bottke2002,
  Weissman2002}. Short--period comets are most likely to orginate from
the Kuiper belt, a reservoir of icy bodies outside the orbit of
Neptune \citep{Levison1997} where their orbits get disturbed as a
result of gravitational perturbations with the giant planets. Entering
the inner Solar System, comets become active through sublimation of
surface volatiles and produce comae and tails. The activity lifetime
of short--period comets \citep[${\sim}$12000~yrs,][]{Levison1997} is
significantly shorter than their dynamical lifetime in near--Earth
space \citep[${\sim}10^7$~yrs,][]{Morbidelli1998}. Hence, it is likely
that the NEO population includes a significant number of
asteroid--like extinct or dormant comets, which have finally or at
least temporarily, ceased being active \citep{Weissman2002}. One
example of a comet that appears to have ceased activity and has become
a dormant or extinct comet is
107P/Wilson--Harrington. Wilson--Harrington was discovered in 1949 as
an active comet, was subsequently lost and re--discovered in 1979 as
NEO (4015) 1979 VA and confirmed as Wilson--Harrington in 1992,
lacking any trace of cometary activity \citep{Bowell1992,
  Fernandez1997}. Vice versa, objects that were originally discovered
as asteroids are later occasionally reclassified after activity was
detected in optical follow--up observations \citep[e.g., see][and
other IAU Circulars]{Warner2005}. Usually, activity is discovered in
such cases a few weeks or months after the discovery of the object
itself.

NEO (3552) Don Quixote was discovered in 1983 as an asteroid, although its
orbit, having a period of 8.68 yrs and a Tisserand parameter with
respect to Jupiter of $T_J=2.313$, resembles very much the orbit of a
typical short--period comet
\citep[e.g.,][]{Hahn1985}. \citet{Veeder1989} obtained
thermal--infrared observations of Don Quixote and used a thermal model
to derive a diameter of 18.7 km and a geometric $V$--band albedo of
0.02, which makes Don Quixote the 3rd--largest known NEO after (1036)
Ganymed and (433) Eros. The low albedo, which agrees well with the
classification of Don Quixote as a D--type asteroid
\citep{Hartmann1987, Binzel2004}, is typical for cometary nuclei
\citep{Lamy2004}. Dynamical simulations by \citet{Bottke2002} predict
a 100\% probability of a short--period comet origin of Don Quixote,
which is the highest probability for a cometary origin among all known
NEOs. In summary, Don Quixote is one of the prime candidates among the
known NEOs for having a cometary origin. Since it has never been
reported to show any sign of activity, it was believed to be an
extinct or dormant comet \citep{Weissman1989, Weissman2002}.

\section{Observations and Data Reduction}

\subsection{Spitzer IRAC Observations}
\label{lbl:irac_observations}

\begin{figure}
\epsscale{1.0}
\plotone{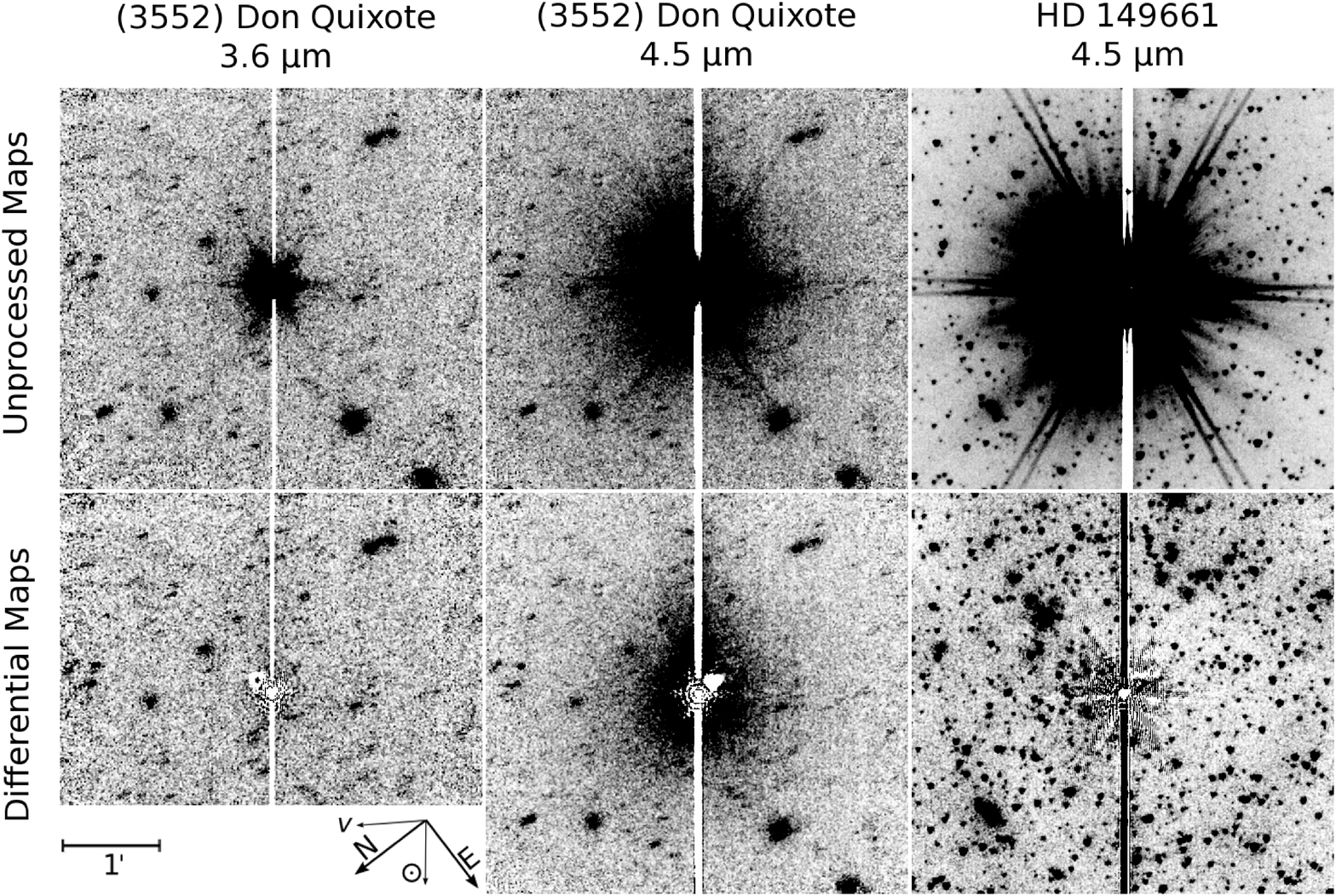}
\caption{Inverted unprocessed (top row) and PSF--subtracted
  differential (bottom row) Spitzer IRAC maps of (3552) Don Quixote at
  3.6~\micron\ (left) and 4.5~\micron\ (center), and a 4.5 \micron\
  map of the saturated calibration star HD149661 (right) for
  comparison. The white bars are image artifacts, caused by the
  well--known ``column pull--down'' effect \citep{IRAChandbook, SOM},
  observed in IRAC channel 1 and 2 mosaics. The white fringes and
  triangular areas in the differential mosaics are the result of a
  misalignment of the model and object point--spread functions (PSFs)
  during the subtraction, and well--understood ghost images of the
  overexposed target, respectively. The differential 4.5~\micron\ map
  of Don Quixote clearly shows a diffuse, elongated feature centered
  on the saturated object that is not visible in the 3.6~\micron\
  map. The feature is extended in the anti--solar direction, as
  indicated in the bottom left map ($v$ indicates the velocity vector
  of the object). The differential mosaic of HD149661 does not show
  any extended emission.  HD149661 is 3.8 mag brighter than Don
  Quixote at the time of observation; hence, any saturation effect
  producing the radially symmetric extended emission around Don
  Quixote would lead to the same effect in
  HD149661. \label{fig:images}}
\end{figure}

Don Quixote was observed by the Infrared Array Camera
\citep[IRAC,][]{Fazio2004} onboard the {\it Spitzer} Space Telescope
\citep{Werner2004} on August 22, 2009, at 19:48 UT. The observations
at 3.6 and 4.5~\micron\ were taken within the ExploreNEOs program
\citep{Trilling2010}, which performed thermal--infrared observations of
$\sim$600 NEOs. At the time of the Spitzer observations, which took
place 18 days prior to Don Quixote's perihelion passage, the target
had a heliocentric distance of 1.23 AU, a solar phase angle of
55\degr, and was 0.55 AU from Spitzer.

The observations (Astronomical Observation Request, AOR, 32690176)
consist of 9 individual 12 sec frames in each band.  The AOR used the
``Moving Cluster'' mode and changed the telescope pointing in such a
way that the source was placed alternately on the 3.6 and 4.5~\micron\
arrays, in order to obtain a nearly--simultaneous dataset in both
bands \citep{Trilling2010}. This mode also maximizes the relative
motion of the asteroid across the field in each band, making it easier
to reject background emission by combination of the individual frames.
Additionally, the pointing was offset relative to the predicted object
position for each frame differently in order to provide dithering for
each band. All frames share the same orientation in the plane of the
sky: the Spitzer spacecraft and IRAC instrument designs require the
Sun to be positioned below the detector array in order to provide
proper shielding from sunlight. Consequentially, the Spitzer--Sun
vector coincides with the pixel array columns.

Mosaics at 3.6 and 4.5~\micron\ were constructed using the
IRACproc software \citep{Schuster2006}, which aligns and combines the
individual frames in each band in the rest frame of Don Quixote, based
on its projected motion. Cosmic rays are filtered and rejected from
the mosaic. Since Don Quixote moved only a few arc seconds during the
course of the observations, the background objects in the field are
not totally rejected but appear as trails in the mosaics.  In the
following, these maps are referred to as ``unprocessed maps'' in
the sense of the further analysis; the maps are shown in the top row
of Figure \ref{fig:images}.

We had selected the integration time of the observations to provide
adequate signal--to--noise ratio within the linear range of the
detectors.  However, due to a failure in the proper retrieval of the
object geometry during the observation planning, the integration time
was overestimated and the observations were found to be saturated in
both the 3.6 and 4.5~\micron\ bands.  In order to estimate the flux of
the point--like source, we apply a technique of subtracting a
calibrated point--spread function (PSF) from the data. By aligning and
scaling a model PSF to the observations, using a
least--squares--method that minimizes the residual, the resulting
scaling factor provides a measure of the object's flux density.  Only
the non--saturated PSF wings and diffraction spikes are used in the
scaling, the saturated regions as well as the column pull--down
regions are masked off in the process \citep{Marengo2009}. Since Don
Quixote was observed in the post--cryo or ``Warm Spitzer'' mission
phase, we use a model PSF that was determined from warm mission
observations of calibration stars with a range of flux densities
\citep{Hora2012, Marengo2013}.

\begin{figure}
\epsscale{0.8}
\plotone{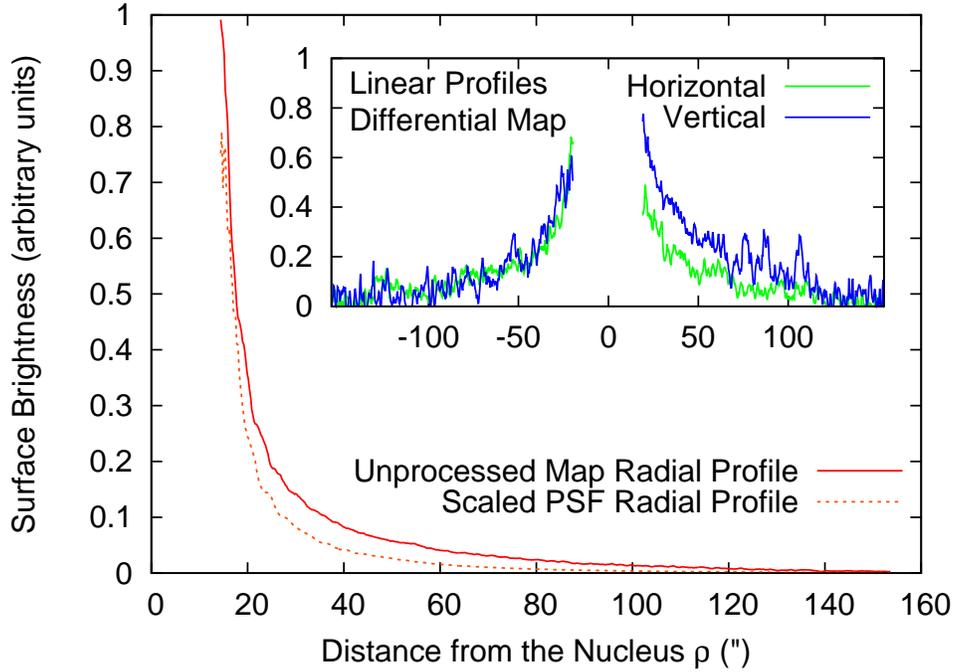}
\caption{Radial average brightness profiles of the unprocessed
    4.5~\micron\ map (red continuous line) and the PSF scaled to fit
    the nucleus of Don Quixote (orange dotted line). The radial
    profiles were produced by averaging the pixel values in annuli
    with a width of 1 pixel, centered on the object center. Each point
    on the profile equals the median value derived from the annulus
    with the respective distance from the center. The space between
    the two lines describes the brightness of the extended
    emission. This plot illustrates the proper scaling of the PSF that
    has been used in the production of the differential image. The
    inlay shows the horizontal (green line) and vertical (blue line)
    linear brightness profiles through the center of Don Quixote,
    generated from the differential 4.5~\micron\ image. Each profile
    represents the median of a 40--pixel wide strip centered on the
    respective axis (column--pull--down areas are masked).  Axis
    definitions are the same as in the outer plot. The vicinity of the
    nucleus is dominated by noise caused by image artifacts and
    therefore not shown here. The agreement between the horizontal and
    vertical profiles is good below the nucleus (negative vertical
    distance). Above the nucleus (positive vertical distance) both
    profiles deviate, which is due to tail emission (see Figure
    \ref{fig:tail}).\label{fig:diff_radplots}}
\end{figure}

When this technique is applied to Don Quixote the derived flux density
at 4.5~\micron\ is much higher than expected, giving an
unrealistically low albedo value using our default thermal modeling
pipeline.  A check of the 4.5~\micron\ ``differential map'', which is
created by subtraction of the fitted PSF from the unprocessed map,
reveals a remnant emission surrounding the core of the object (see
Figure \ref{fig:images}, bottom row). This ``extended emission'' is
not subtracted by the fitted PSF, which means that it does not
originate from a point--like but an extended source. Figure
\ref{fig:diff_radplots} compares the radial brightness profile of the
unprocessed 4.5~\micron\ image with that of the scaled PSF used in the
subtraction process, revealing a discrepancy between the two caused by
the extended emission. Horizontal and vertical inear brightness
profiles through the nucleus also show an excess brightness towards
the top of the image as shown in Figure \ref{fig:images}, which hints
towards the existence of a tail. The same PSF subtraction method
applied to the 3.6~\micron\ map does not show such remnant
emission. Artifacts that are caused by the saturation of the object's
center and a misalignment of the model and image PSFs appear as bright
areas in the bottom row of Figure \ref{fig:images}.

We further investigate the emission from the point--like nucleus in
Section \ref{lbl:nucleus_photometry} and the extended emission in
Section \ref{lbl:halotail_photometry}, separately.

\subsection{Additional Observation Data}

\begin{figure}
\epsscale{0.8}
\plotone{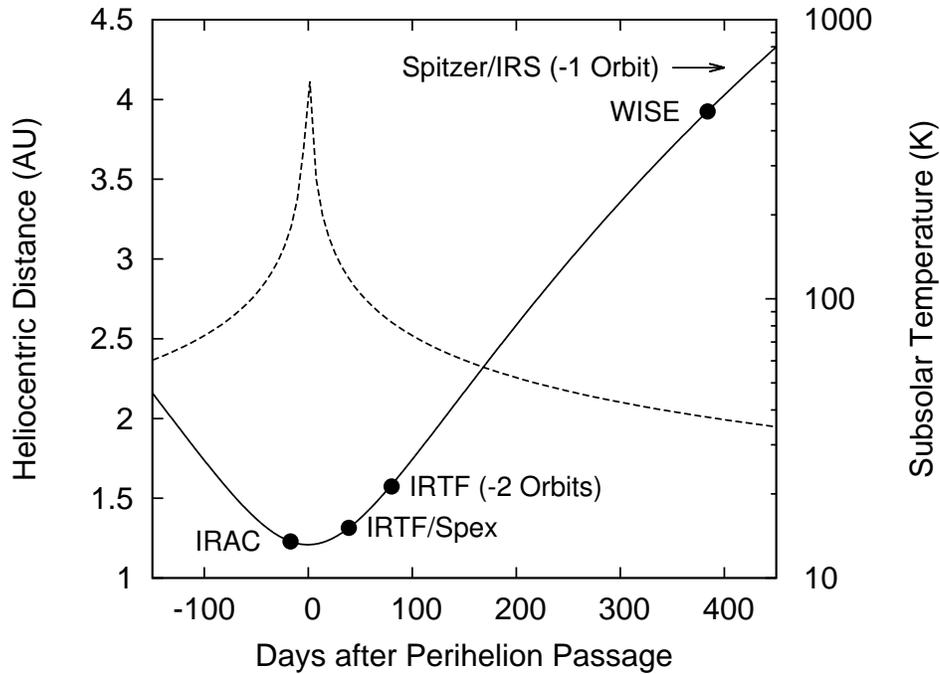}
\caption{Don Quixote's heliocentric distance (continuous line) and
  subsolar temperature (dashed line) as a function of the number of
  days after its perihelion passage. The dots indicate the relative
  position in its orbits at the time of the individual observations
  listed in Table \ref{tbl:fluxes}. Observations that were performed
  in a different orbit cycle are marked. The subsolar temperature is
  based on the thermal model fit performed in Section
  \ref{lbl:thermalmodeling}. Spitzer IRS observations are located
  outside the plot and are indicated by an
  arrow.\label{fig:observations}}
\end{figure}

We have searched the literature for previous observations of Don
Quixote in various wavelength regions to compare our findings with. We
did not succeed in finding reliable optical photometry that is useful
for our purposes. We found useful thermal--infrared data from the literature
as listed in Table \ref{tbl:fluxes} and discussed below. The
heliocentric distance of Don Quixote at the time of the individual
observations ist illustrated in Figure \ref{fig:observations}.

\begin{table}
\begin{center}
\small
\caption{Don Quixote Observations.\label{tbl:fluxes}}
\begin{tabular}{rcccccccl}
  \tableline\tableline
  Observatory & Date \& Time & $\lambda$ & $r$ & $\Delta$ & $\alpha$ & F
  & $\sigma_F$ & Ref. \\
  & (UT) & (\micron) & (AU) & (AU) & (\degr) & (mJy)
  & (mJy) & \\ 
  \tableline
  Spitzer/IRAC\tablenotemark{\star} & 09-08-22 19:48 & 3.6 & 1.229 & 0.550 & 55.4 & 210 &
  10 & 1\\
  Spitzer/IRAC\tablenotemark{\star} & 09-08-22 19:48 & 4.5 & 1.229 & 0.550 & 55.4 & 970 & 50
  & 1\\
  Spitzer/IRS Peakup & 04-03-23 04:40 & 16.0 & 6.910 & 6.494 & 7.8 & 6.30 & 0.33 & 1 \\
  Spitzer/IRS Spectrum & 04-03-23 04:40 & S & 6.910 & 6.494 & 7.8 & -- & -- & 1 \\
  IRTF/SpeX & 09-10-18 05:49 & S & 1.314 & 0.303 & 15.5 & -- & -- & 3 \\ 
  IRTF & 83-10-13 09:21 & 10.1 & 1.574 & 0.664 & 23.1 & 9000\tablenotemark{\dagger} & 100 & 2 \\
  IRTF & 83-10-13 10:04 & 10.1 & 1.575 & 0.665 & 23.1 & 7200\tablenotemark{\dagger} & 100 & 2 \\
  WISE & 10-09-27 13:44 & 3.4 & 3.924 & 3.818 & 14.8 & 0 & 0.1 & 4 \\
  WISE & 10-09-27 13:44 & 4.6 & 3.924 & 3.818 & 14.8 & 0 & 0.2 & 4 \\
  WISE & 10-09-27 13:44 & 11.6 & 3.924 & 3.818 & 14.8 & 28 & 6 & 4 \\
  WISE & 10-09-27 16:54 & 3.4  & 3.925 & 3.817 & 14.8 & 0 & 0.1 & 4 \\
  WISE & 10-09-27 16:54 & 4.6 & 3.925 & 3.817 & 14.8 &  0 & 0.2 & 4 \\
  WISE & 10-09-27 16:54 & 11.6 & 3.925 & 3.817 & 14.8 & 36 & 7 & 4 \\
  WISE & 10-09-28 07:11 & 3.4 & 3.928 & 3.813 & 14.8 & 0 & 0.1 & 4 \\
  WISE & 10-09-28 07:11 & 4.6 & 3.928 & 3.813 & 14.8 & 0.1 & 0.1 & 4 \\
  WISE & 10-09-28 07:11 & 11.6 & 3.928 & 3.813 & 14.8 & 47 & 5 & 4 \\
  WISE & 10-09-28 10:22 & 3.4 & 3.922 & 3.812 & 14.8 & 0 & 0.1 & 4 \\
  WISE & 10-09-28 10:22 & 4.6 & 3.922 & 3.812 & 14.8 & 0 & 0.1 & 4 \\
  WISE & 10-09-28 10:22 & 11.6 & 3.922 & 3.812 & 14.8 & 39 & 8 & 4 \\
\end{tabular}
\tablecomments{$F=0$ refers to a non--detection of the object; the
  respective flux density uncertainty then gives the 95\% confidence
  upper limit flux density. The meanings of the columns are: Date \&
  Time: observation midtimes (YY-MM-DD HH:MM),
  $\lambda$: monochromatic wavelength (``S'' in case of spectroscopic
  observations), $r$: heliocentric distance, $\Delta$: distance from
  the observer, $\alpha$: solar phase angle, $F$: measured flux
  density (not color--corrected, if not mentioned otherwise),
  $\sigma_F$: 1$\sigma$ uncertainty of the measured flux;}
\tablenotetext{\star}{Spitzer/IRAC flux densities of Don Quixote refer
  to the thermal--infrared emission of the nucleus only;}
\tablenotetext{\dagger}{flux densities from \citet{Veeder1989} are
  color--corrected;} \tablerefs{(1) this work; (2) \citet{Veeder1989};
  (3) \citet{Thomas2013}; (4) WISE data as extracted from the WISE
  3--Band Cryo Known Solar System Object Possible Association List
  (see Section \ref{lbl:wise});}
\end{center}
\end{table}

\subsubsection{WISE Observations}
\label{lbl:wise}

The Minor Planet Center reports four observations of the ``Wide--Field
Infrared Survey Explorer'' \citep[WISE,][]{Wright2010} of Don Quixote
in September 2010 during the ``3--band cryogenic'' phase of the
mission. The observations took place 410 days after the perihelion
passage during the same orbit as the IRAC observations. The measured flux
densities were accessed via the NASA/IPAC Infrared Science
Archive\footnote{http://irsa.ipac.caltech.edu/Missions/wise.html} and
extracted from the ``WISE 3--Band Cryo Known Solar System Object
Possible Association List''. The reported magnitudes were converted
into flux density units using the zeropoint magnitudes reported in
\citet{Wright2010}; the flux densities are listed in Table
\ref{tbl:fluxes}. See \citet{Mainzer2011}, \citet{Mainzer2012}, and
references therein for a full discussion of asteroid observations with
WISE. As it turned out, Don Quixote was too faint to be clearly
detected in most of the 3.5 and 4.6~\micron\ measurements; most of the
data represent 2$\sigma$ upper limit flux densities. Low
signal--to--noise observations are available at 11.6~\micron. 

\subsubsection{IRTF Photometry}

Don Quixote was observed by \citet{Veeder1989} using the NASA Infrared
Telescope Facility (IRTF). They report two $N$--band magnitudes
measured on October 13, 1983, which were here converted into flux density
units using a calibration spectrum of Vega \citep{Rieke2008} and are
listed in Table \ref{tbl:fluxes}. The observations of
\citet{Veeder1989} took place 80 days after its perihelion, two
orbits earlier than our IRAC observations.

\subsubsection{IRTF SpeX Spectroscopy}

Spectroscopic observations of Don Quixote have been obtained in the
wavelength range 0.6--2.6~\micron\ using the SpeX instrument
\citep{Thomas2013}. SpeX \citep{Rayner2003} is a medium resolution
spectrograph and imager unit at the IRTF. The SpeX spectrum was
obtained on October 18, 2009, 40 days after its perihelion passage in
the same orbit of Don Quixote as the IRAC observations. Don Quixote
and the solar standard star Landolt 113-276 were observed
close in time. The data were reduced using SpeXtool
\citep{Cushing2004} and the telluric atmosphere correction was done
using the ATRAN model atmosphere
\citep[e.g.,][]{Lord1992,Rivkin2004}. The spectrum discussed in
  Section \ref{lbl:specdata} was produced by division of the measured
spectrum of Don Quixote by that of the solar analog star, resulting in
a measure of the reflectance of the object's surface. For more
information on the processing of the spectrum see \citet{Thomas2013}.

\subsubsection{IRS Peakup Imaging and Spectroscopy}

Don Quixote was also observed with the Infrared Spectrograph
\citep[IRS,][]{Houck2004} onboard the Spitzer Space Telescope. The
observations (AOR 4869888) took place on March 23, 2004, 3.2 yrs after
perihelion when Don Quixote was 6.9 AU from the Sun, one orbit earlier
than the Spitzer IRAC observation. The IRS observations were made
using only the long wavelength, low resolution (LL) modules (LL2, 14.2
to 21.7~\micron; LL2, 19.5 to 38.0~\micron), including IRS Peakup
observations, from which a flux density could be derived (see Table
\ref{tbl:fluxes}). All data have been processed through the standard
point--source pipeline (V18.18) by the Spitzer Science Center to
produce Basic Calibrated Data (BCD).  During the observations, the
object was nodded along the slit.  We subtract the BCD frames of the
two nod positions for each module to remove background flux.  The data
are then extracted to 1--D spectra by summing data within each
constant wavelength polygon and apply the wavelength calibrations
supplied by the Spitzer Science Center (SSC).  We developed custom
routines for IRS spectral reduction.  Therefore, rather than applying
the absolute flux calibration supplied by the SSC (which is valid only
for their exact extraction parameters), we construct our own absolute
and relative spectral calibration factors from standard stars observed
by IRS throughout the mission.  This method enables the flexibility to
adjust the extraction width to optimize the signal--to--noise ratio.
Multiple cycles and nod positions for each module are averaged, and
the LL2 and LL1 spectra are scaled to each other by matching flux in
the spectra range of overlap.  Our reduction procedure is described in
more detail by \cite{Emery2006}.

\section{Spitzer/IRAC Data Analysis}

\subsection{Emission from the Nucleus}
\label{lbl:nucleus_photometry}

In order to estimate the flux density of Don Quixote's unresolved
nucleus, we fit a calibrated PSF to the unsaturated regions of its
image as described in Section \ref{lbl:irac_observations}. For the
4.5~\micron\ map, the fit is done manually and the scaling of the PSF
iterated until the residuals are minimized in the difference
image. The radial profiles of the unprocessed image and the scaled PSF
are shown in Figure \ref{fig:diff_radplots}. The plot shows that a
significant part of the total image brightness is emission from the
nucleus of Don Quixote. For an annulus with inner and outer radius of
20\arcsec\ and 45\arcsec, respectively, centered on Don Quixote, more
than half of the total brightness is due to emission from the
nucleus. The ratio drops to 30\% at radii larger than 80\arcsec. These
ratios allow for a proper scaling of the PSF that is subtracted from
the unprocessed map. In this work, we disregard any image data within
a radius of 20\arcsec\ around the nucleus of Don Quixote due to
dominant image artifacts in this area, caused by the saturation and
PSF subtraction.  The differential maps are shown in the bottom row of
Figure \ref{fig:images}. The derived flux densities, as listed in
Table \ref{tbl:fluxes}, are those of the point--like nucleus of the
object.  The same method has been applied by \citet{Mommert2013} on
two other NEOs with saturated IRAC observations, not revealing similar
features. For bright calibration stars, this technique achieves a
typical calibration accuracy on the order of 1\% \citep{Marengo2009}.
In order to account for the increased calibration uncertainty of Don
Quixote due to the relatively fainter core compared to the calibration
standards and the difficulty caused by the extended coma emission, we
add an additional 5\% uncertainty in quadrature to the measured flux
density uncertainties.

\subsection{Emission from the Coma and the Tail}
\label{lbl:halotail_photometry}

\begin{figure}
\epsscale{0.8}
\plotone{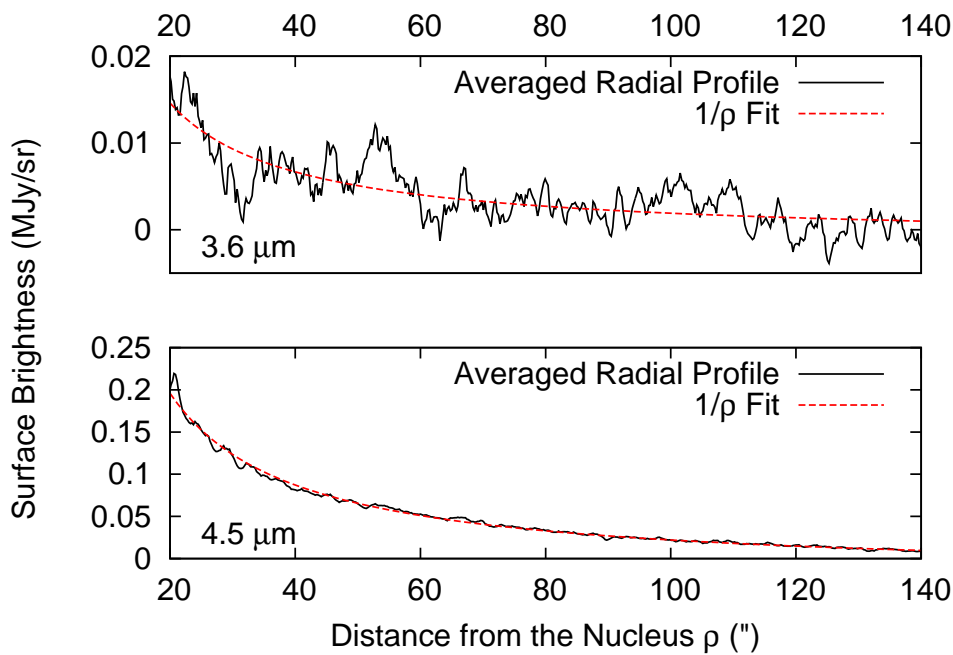}
\caption{Radial average brightness profiles of the differential
  3.6~\micron\ (top) and 4.5~\micron\ (bottom) maps, produced using
  the same method as in Figure \ref{fig:diff_radplots}.  Both radial
  profiles were fitted over the whole range plotted here. The
  3.6~\micron\ radial profile suffers from a low signal--to--noise
  ratio of the emission. Nevertheless, it shows a weak trend in
  surface brightness versus radial distance. Both profiles have been
  fitted using a 1/$\rho$ relation (red dashed lines), which is the
  behavior expected for outgassing phenomena.  The 4.5~\micron\ radial
  profile agrees well with this behavior, whereas the 3.6~\micron\
  profile agrees only coarsely, due to the weak signal. Note the
  different scales on the vertical axes of both
  panels.\label{fig:radialfit}}
\end{figure}

A detailed inspection of the IRAC maps after the PSF subtraction
revealed extended emission in the form of a mostly radial symmetric
coma--like structure in the 4.5 \micron\ map (Figure \ref{fig:images},
bottom row).  In contrast, the 3.6 \micron\ map shows no sign of a
diffuse source component. The extended emission at 4.5 \micron\ also
shows a tail--like elongation towards the top of the map, pointing
away from the direction towards the Sun.

\begin{figure}
\epsscale{0.5}
\plotone{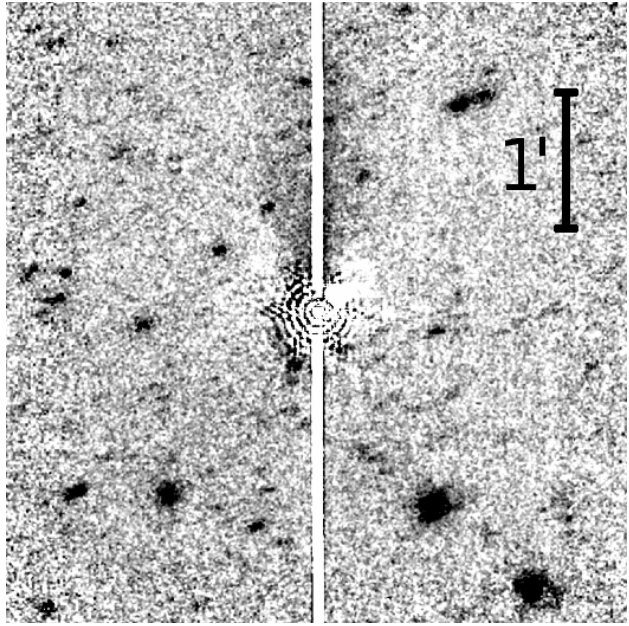}
\caption{Don Quixote's tail. This map shows the differential
    4.5~\micron\ map of Don Quixote from which the fitted $1/\rho$
    radial profile shown in Figure \ref{fig:radialfit} has
    been subtracted. The resulting map clearly shows the tail of Don
    Quixote with a length of ${\sim}2$\arcmin. The white concentric
    rings around the object center are image artifacts from the PSF
    subtraction.\label{fig:tail}}
\end{figure}

In order to derive a quantitative estimate of the extended emission,
we radially average the PSF--subtracted maps using a median
algorithm. Areas affected by the ``column--pulldown effect''
\citep[see Figure \ref{fig:images} and the][]{SOM} are excluded from
the averaging. Figure \ref{fig:radialfit} shows the results of the
fitting in both bands. The extended emission in the 4.5~\micron\ map
clearly follows a $1/\rho$ profile, where $\rho$ is the angular
distance from the object's center. This is the radial profile
predicted for free expansion of material from a nucleus, e.g., from
sublimating ices, and is characteristic of cometary comae
\citep{Jewitt1987}. Subtracting the $1/\rho$ profile from the
differential map improves the visibility of the faint cometary tail
with a length of ${\sim}2$\arcmin\ (see Figure \ref{fig:tail}), which
points away from Sun. Note that the direction to the Sun coincides
with the pixel array columns as a result of the Spitzer spacecraft
design. The radial profile of the 3.6~\micron\ map coarsely agrees
with a 1/$\rho$ profile (Figure \ref{fig:radialfit}), despite its low
signal--to--noise ratio. We measure the total flux densities of the
extended emission in both bands by integrating over the fitted
profiles and subtracting the background, yielding $6\pm10$~mJy and
$65\pm10$~mJy at 3.6 and 4.5~\micron, respectively. Due to the large
uncertainty in the 3.6~\micron\ flux density we adopt the derived
value of 6~mJy as an upper flux density limit. The background level
and its uncertainty were measured as the median and standard
deviation, respectively, in four different areas of both maps that are
unaffected by background sources. The measured flux densities were
aperture corrected using the IRAC surface brightness correction
factors \citep{IRAChandbook}.

The derivation of the intensity of the emission from the tail suffers
from the low signal and strong noise in the differential map (Figure
\ref{fig:tail}), precluding a quantitative analysis of the emission
from the tail.

\subsection{Thermal Modeling of the Nucleus}
\label{lbl:thermalmodeling}

\begin{figure}
\epsscale{0.8}
\plotone{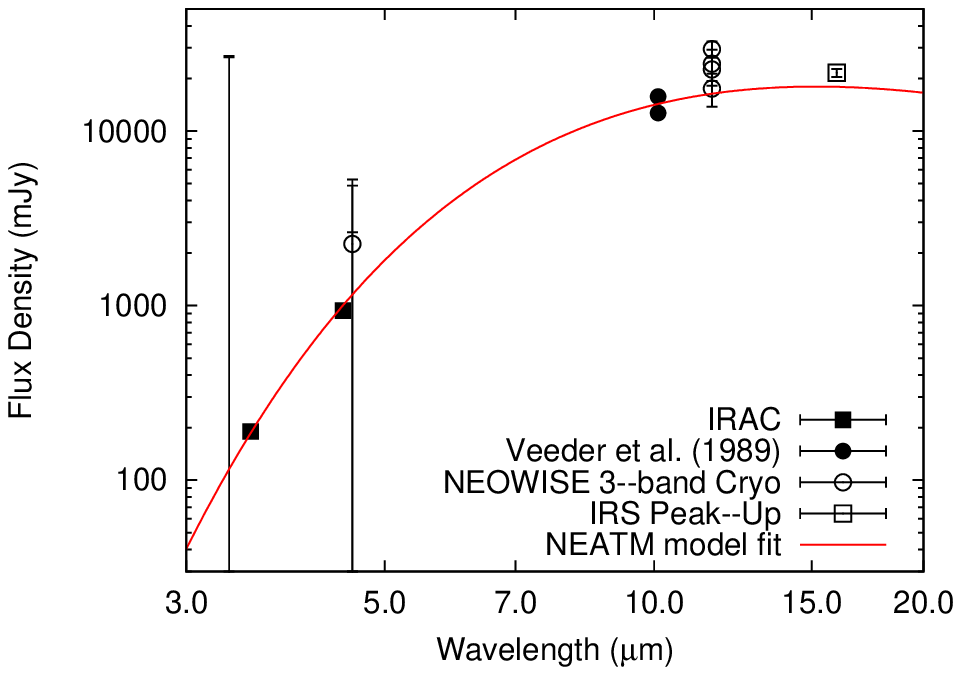}
\caption{Thermal--model fit of Don Quixote's nucleus. The flux
  densities used in the modeling and depicted in the plot are listed
  in Table \ref{tbl:fluxes}. Note that all flux densities shown here
  are color--corrected, corrected for contributions from reflected
  solar light, and normalized: the individual flux densities are
  scaled by the ratio of the best--fit model flux densities for the
  respective epoch and the Spitzer IRAC epoch. Flux uncertainties are
  shown for each datapoint and represent 1$\sigma$ uncertainties. The
  line depicts the best--fit thermal model using the
  NEATM.\label{fig:thermalmodel}}
\end{figure}

We use the Near--Earth Asteroid Thermal Model
\citep[NEATM,][]{Harris1998} to derive the diameter ($d$) and the
geometric $V$--band albedo ($p_V$) of Don Quixote's nucleus. The NEATM
combines thermal infrared and optical data to derive that set of
diameter and albedo that provides the best fit to the measured
spectral energy distribution. The fitting routine uses a variable
beaming parameter \citep[$\eta$,][]{Harris1998} that modulates the color
temperature of the model SED. The NEATM is widely used to
derive the physical properties of asteroids \citep[see,
e.g.,][]{Trilling2010, Mainzer2011}, as well as cometary nuclei
\citep[see, e.g.,][]{Lamy2004}.

We apply the NEATM on the thermal--infrared flux densities listed in
Table \ref{tbl:fluxes} and adopt the $H$ magnitude estimate from the
Minor Planet Center\footnote{http://www.minorplanetcenter.net/}
($H=13.0$~mag with an uncertainty estimate of 0.5~mag). The individual
geometry for each epoch is properly taken into account in the thermal
modeling.  We use an iterative color correction of the IRAC, WISE and
IRS Peakup data, and subtract the contributions of reflected solar
light as explained in \citet{Trilling2010} and \citet{Harris2011}. The
resulting best--fit diameter is 18.4$_{-0.4}^{+0.3}$~km and the albedo
$p_V=0.03^{+0.02}_{-0.01}$, using a best--fit $\eta=0.82\pm0.03$. Our
results agree well with earlier estimates of the physical properties
of Don Quixote \citep[][see also Section
\ref{lbl:intro}]{Veeder1989}. We confirm that Don Quixote is the
third--largest known NEO after (1036) Ganymed
\citep[d=38.5~km,][]{Veeder1989} and (433) Eros
\citep[d=23~km,][]{Harris1999}. The discovery of cometary activity in
Don Quixote makes this object also one of the largest short--period
comets with a measured diameter \citep{Lamy2004}.  Figure
\ref{fig:thermalmodel} shows the good fit of the model spectral energy
distribution to the measured thermal--infrared flux densities.

\section{Discussion}

\subsection{Discussion of the Spitzer IRAC Data}

In the following, we discuss the robustness of our observations
and rule out the possibility of the extended emission to be an image
artifact.

\begin{itemize}

\item {\it The observed emission is not a background object.}
  Inspection of the field of the sky in which Spitzer observed Don
  Quixote in the Digitized Sky Surveys \citep{DSS}, the Two Micron All
  Sky Survey \citep{Skrutskie2006}, as well as in WISE channels
    W1 (3.4~\micron) and W2 (4.6~\micron) of the WISE All--Sky Data
    Release \citep{Cutri2012}, show no extended object bright enough
  to be the source of the observed diffuse emission. The closest
  bright star, HD 22634 with $V=6.7$ mag, is separated from Don
  Quixote at the time of the observation by some 6.3\arcmin, outside
  the field of view.

\item {\it We can rule out stray--light or scattered light as the
    source of the emission.} The stray--light and scattered--light
  behavior of Spitzer's IRAC instrument is well--understood \citep[see
  the][]{IRAChandbook, SOM}. At the time of the IRAC observations no
  sufficiently bright background sources were present in the
  stray--light avoidance zones of either IRAC channel. In order to
  rule out the possibility of a contamination by stray--light or ghost
  images entirely, we subtract individually normalized PSFs from each
  of the 9 frames taken in the 4.5~\micron\ band that were used in the
  generation of the unprocessed map. The position of the extended
  emission is centered on the object in all individual frames; the
  extent and intensity of the emission is equal in all frames, as
  well. In the case of a contamination by stray--light, the dithering
  would force significant variations in the intensity and position of
  the resulting ghost image. Hence, we can confidently rule out the
  possibility of the extended emission being a ghost image.

\item {\it The extended emission is not caused by latency effects.}
  The PSF subtraction from the individual frames shows the extended
  emission to be centered around the object in each frame. This would
  not be the case if the emission were an image artifact caused by
  latency effects, i.e., left--over charge in the pixel wells from
  previous integrations, given the dithering between the individual
  frames.

\item {\it The extended emission is not an image artifact caused by
    the saturation of the object.} We can rule out the possibility of
  the coma being an artifact caused by the saturation of the mosaics,
  since the detector behavior is
  well--characterised \citep{IRAChandbook, SOM}. For comparison
  reasons, we have examined observations of stars with a wide range of
  brightness, many of which are saturated, but none of which show
  extended emission. As an example, we show a saturated image of
  calibration star HD 149961 in Figure \ref{fig:images}, rightmost
  column. HD 149961 is significantly brighter ($\Delta m = 3.8$ mag)
  than Don Quixote at the time of its observation. The image does not
  show any sign of radially symmetric extended emission.

\item {\it The extended emission is not an image artifact introduced
    by the subtraction of the PSF.} We have applied the PSF
  subtraction technique \citep{Marengo2009} to images of calibration
  stars taken during the cryogenic and ``warm'' mission phases of
  Spitzer, using the respective PSF, and found no equivalent to the
  extended emission observed in Don Quixote. Improper scaling of the
  PSF can lead to residuals in the differential image. In that case,
  however, residuals of the spikes would be visible, which form the
  brightest parts of the wings of the PSF. Improper aligning of the
  PSF with respect to the object leads to artifacts that do not have the
  radial symmetric nature of the extended emission observed in Don
  Quixote.
\end{itemize}

\subsection{Constraining the Nature of the Emission}

The nature of the extended coma--like emission is constrained by the
ratio of the infrared flux densities, $F_{4.5}/F_{3.6}$, which has a
value of ${\geq}9.2$, using the flux density upper limit at
3.6~\micron\ and taking into account the 1$\sigma$ uncertainty of the
4.5~\micron\ flux density (see Section
\ref{lbl:halotail_photometry}). Based on a model for cometary dust
\citep{Kelley2009, Reach2013}, the expected ratio of $F_{4.5}/F_{3.6}$
for thermal emission and reflected sunlight from the dust is less than
5 for a comet at 1.23 AU from the Sun. We are confident that the
source of the higher than expected 4.5~\micron\ flux density is
molecular band emission of CO (at 4.7~\micron) or CO$_2$ (at
4.3~\micron\ and 15.0~\micron) that are stimulated by
photo--dissociation and fall well within the IRAC 4.5~\micron\
bandpass. Both molecular bands have been observed in many comets, with
CO$_2$ typically dominating for comets in the inner Solar System
\citep{Ootsubo2012, Reach2013}. Hence, we focus on a CO$_2$ origin of
the observed emission. CO$_2$ molecular band emission explains the
lack of extended emission in the 3.6~\micron\ band. If the detected
emission at 3.6~\micron\ is real, it is most likely reflected solar
light from dust particles that are launched from the surface by the
CO$_2$ gas drag, according to the cometary dust model
\citep{Reach2013}.

The upper limit nature of both flux density measurements of the tail
(see Section \ref{lbl:halotail_photometry}) precludes the use of the
$F_{4.5}/F_{3.6}$ ratio as an indicator for the nature of the
emission. We suppose the nature of the tail emission to be either
molecular band emission as in the coma or solar light that is
reflected from dust particles. We investigate the possibility that the
tail emission is solely caused by CO$_2$ band emission.  The length of
the tail shown in Figure \ref{fig:tail} is ${\sim}2$\arcmin, which
equals ${\sim}48000$~km at the distance of Don Quixote. Assuming an
expansion velocity of the gas of $0.72$~km s$^{-1}$
\citep[0.8~km~s$^{-1}$\ $\times\ r^{-0.5}$ with
$r=1.23$~AU,][]{Ootsubo2012}, the average lifetime of the particles is
required to be ${\geq}0.77$ days to be able to explain the observed
tail, which is well within the lifetime for dissociation by sunlight
of CO$_2$ \citep[8.6 days, data from][normalized to $r=1.23$~AU
assuming an inverse--square relationship between the lifetime and the
heliocentric distance]{AHearn1995}. Hence, we cannot exclude the
possibility that the tail emission is molecular band emission.

\subsubsection{Gas and Dust Production Rates}
\label{lbl:productionrates}

We estimate the gas and dust production rates from the measured 4.5
and 3.6~\micron\ extended emission flux densities, respectively,
assuming (1) the 3.6~\micron\ flux density to be purely reflected
solar light from dust grains, and (2) the 4.5~\micron\ flux density to
be dominated by band emission, with a contribution from thermal
emission from dust. Taking into account the uncertain nature of the
measured 3.6~\micron\ flux density, we treat it as a 6~mJy upper
limit.

We adopt the widely used $A f \varrho$--formalism, introduced by
\citet{AHearn1984}, to determine the properties of the dust coma,
based on the assumption that the upper limit flux density at
3.6~\micron\ is solely reflected solar light. $A f \varrho$, measured in
units of cm, is the product of the dust grain bond albedo ($A$), the
filling factor of the grains ($f$), and the linear
radius\footnote{$\varrho(\mathrm{cm}) =7.25\cdot10^7 \times \Delta(\mathrm{AU})
  \times \Theta(\arcsec)$, where $\Theta$ is the angular radius of the
  aperture in which the flux density of the coma was measured in units
  of arc seconds and $\Delta$ is the comet--observer distance in AU.}
of the field of view at the distance of the comet ($\varrho$), and is
hence independent of the characteristics of the observation.
\begin{equation} 
A f \varrho = \frac{(2 \Delta r)^2}{\varrho} \frac{F_c}{F_s},
\end{equation}
where $r$ is the heliocentric distance of the comet in AU, $\Delta$
the distance to the observer in cm, $F_c$ and $F_s$ are the
measured flux density of the coma and the solar light flux density at
1 AU, respectively, in the same band. We use $F_c = 6$~mJy with
  an angular radius of the aperture of 260\arcsec\ and determine
$F_s=5.7\cdot10^{16}$~mJy by integration of the measured solar
spectrum by \citet{Rieke2008} convolved with the spectral response
function of the IRAC 3.6 \micron\ band over its bandwidth. We find $A
f \varrho \leq 4$~cm, which is lower than most other short--period
comets \citep{AHearn1995}. 

\noindent $A f \varrho$ can be converted into a dust production rate 
\begin{equation}
Q_{\mathrm{dust}} = (A f \varrho)\ \frac{2}{3}\frac{\rho_d a
  v_d}{A_p},\label{eqn:dustproduction}
\end{equation}
where $\rho_d$ is the dust density, $a$ the dust grain radius, $v_d$
the escape velocity, and $A_p$ the geometric albedo of the dust
particles, assuming a fixed grain size \citep{Jorda1995,
  Fornasier2013}. We adopt values that are typical for short--period
comets: $v_d=0.72$~km~s$^{-1}$ (using the expansion velocity of
gas\footnote{Since the dust component is probably driven by the
  sublimation of gas, the use of this relation here is justified.}:
0.8~km~s$^{-1}$\ $\times\ r^{-0.5}$ with $r=1.23$~AU,
\citet{Ootsubo2012}), $a\sim15$~\micron\ \citep[average of the range
of particle sizes found for short--period comet
67P/Churyumov-Gerasimenko by][]{Bauer2012}, $\rho_d=1$~g~cm$^{-3}$
\citep{Bauer2012}, and $A_p = 0.15$ \citep{Kelley2009}. We obtain an
upper limit on the dust production rate of
${\leq}1.9$~kg~s$^{-1}$. This estimate is comparable to other
short--period comets \citep[e.g.,][]{Bauer2011}. Note that such a
  low dust production is barely detectable with optical means, as
  discussed in Section \ref{lbl:implications}.

In the next step, we determine the CO$_2$ gas production rate from the
measured 4.5~\micron\ flux density. Firstly, we correct the measured
flux density for the contribution from thermal emission from dust,
based on the results derived above. We determine the contribution of
thermal emission from dust as the integral over the thermal emission
spectrum of dust convolved with the IRAC 4.5~\micron\ spectral
response function. The thermal emission spectrum of dust is described
using a model provided by \citet{Kelley2009}:
\begin{equation}
  F_{\mathrm{therm}} = \frac{(1-\bar{A})}{A(\alpha)} \pi B_\lambda(T) \frac{(A f
    \varrho)}{\Delta^2} \varrho,
  \label{eqn:thermaldust}
\end{equation}
where $\bar{A}\sim0.32$ is the mean bolometric Bond albedo of the dust
\citep{Gehrz1992}, $A(\alpha)$ is the phase angle dependent Bond albedo
\citep[which is assumed to be 0.15 for $\alpha\leq
60$\degr,][]{Kelley2009}, and $B_\lambda(T)$ is the Planck function
 with temperature $T\sim277$~K \citep[$= 306$~K$ \times
1.23^{-0.5}$,][]{Kelley2009}. Given the upper limit nature of the 
3.6~\micron\ flux density measurement, we constrain that part of the
emission at 4.5~\micron\ resulting from molecular band emission to
the range $51 < F < 65$~mJy.

We determine the CO$_2$ production rate based on the single--species
\citet{Haser1957} model, which describes the number density of
molecules, $n$, in a distance $\varrho$ from the nucleus. The Haser model
assumes the coma to be the result of a uniform, spherically symmetric
outflow of molecules from a point--like nucleus at a constant
speed. The emission is caused by the photo--dissociation of the CO$_2$
molecules. The number density (km$^{-3}$) is defined as
\begin{equation}
  n(\varrho) = \frac{Q}{4\pi \varrho^2v}\ \exp(-\varrho/\gamma),
  \label{eqn:gas_numberdensity}
\end{equation}
where $Q$ is the production rate (s$^{-1}$), $v$ the radial outflow
velocity (km~s$^{-1}$), and $\gamma = \tau v$, the scale length (km),
which is the product of the photo--dissociation lifetime of CO$_2$, $\tau$
(in s$^{-1}$), and the outflow velocity. We adopt the expansion velocity of
gas at the heliocentric distance of Don Quixote, $v=0.8\times 1.23^{-0.5} =
0.72$~km~s$^{-1}$ \citep{Ootsubo2012} and the
lifetime\footnote{\label{fnt:crovisier}J. Crovisier's molecular database:
  \url{http://lesia.obspm.fr/perso/jacques-crovisier/basemole}} of
CO$_2$, also scaled to the heliocentric distance, $\tau = 5.0\cdot10^5
\mathrm{ s } \times 1.23^2 = 7.4\cdot10^5$~s. In order to derive the
production rate $Q$, the column density $N(\varrho)$ in units of
km$^{-2}$ has to be derived
from the number density by integration along the line of sight
\citep[see, e.g.,][]{Helbert2003}, assuming the coma to be optically
thin. The column density is related to the measured flux density $F$
in units of W~m$^{-2}$~\micron$^{-1}$ via 
\begin{equation}
  N(\varrho) = \frac{4\pi F \cdot 10^{-9}}{Q (hc/\lambda) g \pi
    \varrho^2} \Delta^2,
  \label{eqn:gasproduction}
\end{equation}
where $h$ is Planck's constant, $c$ the velocity of light in vacuum,
$\lambda=4.26$~\micron\ the center wavelength of the CO$_2$ emission band, and $g$
the fluorescence efficiency of this band
($g=2.6\cdot10^{-3}$~s$^{-1}$, see Footnote \ref{fnt:crovisier}). The
factor $10^{-9}$ stems from the conversion from km to \micron; 
  the aperture size for the integration of the 4.5~\micron\ flux
  density is 200\arcsec. Solving
this equation yields a CO$_2$ production rate of
$Q=(1.1\pm0.1)\cdot10^{26}$~molecules~s$^{-1}$, which is low but
comparable to other short--period comets that exhibit CO$_2$ emission
at comparable heliocentric distances \citep{Ootsubo2012}.

\subsection{Constraints from Additional Data}

\paragraph{Imaging data.} 

The IRTF flux density measurements from \citet{Veeder1989} have
high signal--to--noise ratios (see Table \ref{tbl:fluxes}) that
might be sufficient to detect emission from a possible coma. However,
the two measured flux densities deviate significantly, rising doubts
about the actual accuracy of the observations and precluding further
analysis.

The low signal--to--noise ratio of the WISE observations (see Table
\ref{tbl:fluxes}) precludes a search for extended emission in the
image data. In the 3.4 and 4.6~\micron\ bands of WISE, which are
important for the confirmation of CO$_2$ band emission, Don Quixote is
barely detected and only upper limit flux densities are available in
all but one case. 

The bandwidth of the IRS Peakup observations (13.0--18.5~\micron)
covers an additional CO$_2$ molecular emission band at 15.0~\micron,
which enables the search for such emission in these data. We use a
method similar to the analysis of the Spitzer IRAC observations (see
Section \ref{lbl:halotail_photometry}). A comparison of the radial
profile of Don Quixote's image to that of a calibration star observed
near in time shows no significant differences, arguing against
cometary activity. In a different approach we subtract a fitted PSF
that has been modeled from a calibration star observed near in time to
Don Quixote. From the residual 3$\sigma$ flux density upper limit
(0.57~mJy) we derive an upper limit on the CO$_2$ gas production rate
in the same way as in Section \ref{lbl:productionrates}, assuming that
all of the residual emission is molecular band emission from
CO$_2$. From Equations \ref{eqn:gas_numberdensity} and
\ref{eqn:gasproduction} we derive $Q_{\mathrm{IRS}} \leq
13\cdot10^{26}$~molecules~s$^{-1}$ as a 3$\sigma$ upper limit, using
$g=8.2\cdot10^{-5}$~s$^{-1}$ (see Footnote \ref{fnt:crovisier}) and
$v=0.8\times 6.91^{-0.5} = 0.30$~km~s$^{-1}$ \citep{Ootsubo2012}. This upper limit is
significantly higher than the gas production range derived in Section
\ref{lbl:productionrates} but is still comparable to CO$_2$ production
rates of other short--period comets \citep{Ootsubo2012}. Note that
this upper limit estimation does not unambiguosly prove the existence
of activity in the IRS Peakup observations.

\paragraph{Spectroscopic data.}
\label{lbl:specdata}

\begin{figure}
\epsscale{0.8}
\plotone{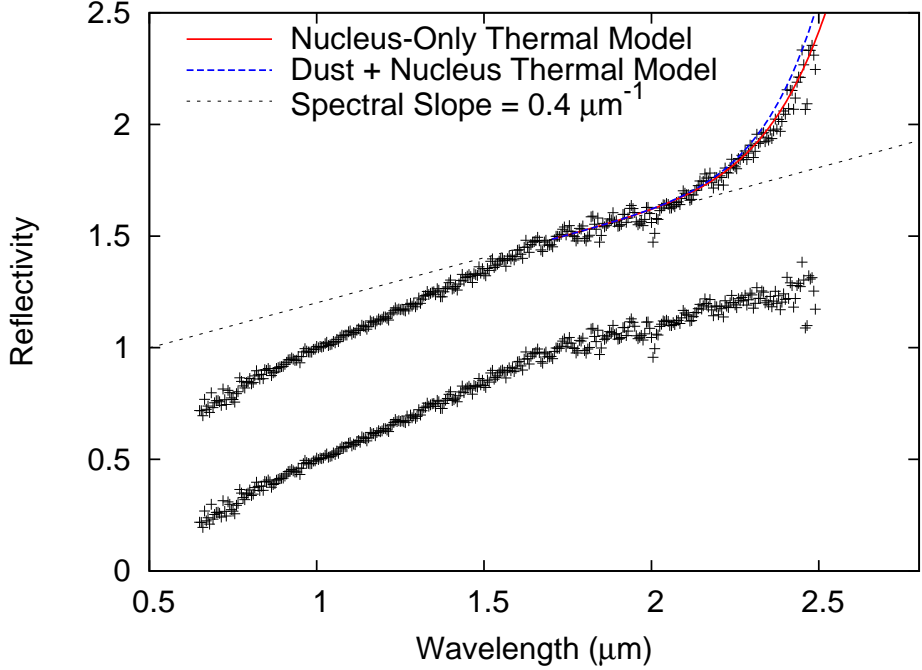}
\caption{IRTF SpeX spectrum of Don Quixote (black crosses, top) with
  different emission models. The continuous red line depicts the
  predicted thermal emission from the nucleus, solely based on the
  best--fit thermal model parameters derived in Section
  \ref{lbl:thermalmodeling}. A possible contribution from thermal
  emission from dust, based on the properties of the dust coma as
  derived from the 3.6~\micron\ flux density, is several orders of
  magnitude fainter than the emission from the nucleus and hence
  barely detectable. In order to derive an upper limit on the dust
  production rate from the spectrum, we apply an amplification factor
  of 100 to the predicted dust emission (dashed blue line). The lower
  plot shows the same spectrum, offset by 0.5 in the ordinate axis,
  from which the thermal tail has been subtracted, assuming the tail
  to be solely caused by the thermal emission from the nucleus
  only. Spectral data are taken from
  \citet{Thomas2013}.\label{fig:thomasspex}}
\end{figure}

Figure \ref{fig:thomasspex} shows Don Quixote's near--infrared
reflectivity as a function of wavelength; the reflectivity is defined
as the ratio of the measured flux density per wavelength and the
spectrum of a solar analog star.  The featureless spectrum has a
spectral slope of ${\sim} 0.7$ for $\lambda{<}1.7$~\micron\ and
${\sim}0.4$ for $\lambda{>}1.7$~\micron, which agrees with the
previous classification of a D--type asteroid \citep{Hartmann1987,
  Binzel2004} and other D--type asteroids \citep[e.g.,][]{Demeo2009,
  Bus2002}. At the long--wavelength end of the spectrum a so--called
``thermal tail'' occurs where the reflectivity seems to increase
dramatically as a result of contributions from thermal emission of the
nucleus and/or the coma dust. We investigate the possible contribution
of thermal emission from dust to the thermal tail.

The dotted line in Figure \ref{fig:thomasspex} presents a linear fit
to the spectrum in the wavelength range 1.7--2.1~\micron\ that is
unaffected by the thermal tail. We compute the thermal emission
spectrum of the nucleus and the dust coma for this wavelength range
using the thermal model of the nucleus (see Section
\ref{lbl:thermalmodeling}) and the model for thermal emission from
dust (Equation \ref{eqn:thermaldust}), respectively. The dust emission
model is based on an aperture size that equals the length of the
spectrograph slit (60\arcsec). The thermal emission from dust is
several orders of magnitude fainter than the emission from the nucleus
and barely affects the shape of the spectrum. The emission from the
nucleus is shown as a continuous red line in Figure \ref{fig:thomasspex}. The
line fits the data points well without using any fitting to the
data. The dashed blue line shows the thermal model of the nucleus combined
with a model for thermal emission from dust that assumes a 100 times
higher dust production rate. The combined emission model still agrees
with the measured spectrum and translates into an upper limit of the
dust production rate of 190~kg~s$^{-1}$. The upper limit is
significantly higher than the value derived from IRAC observations,
but does not prove the presence of activity in the spectral data.

We take a different approach in the analysis of the IRS spectral data,
since the simultaneously acquired IRS Peakup data revealed no clear
evidence for cometary activity. We apply a NEATM fit to the calibrated
spectrum, yielding a diameter of 18.8$\pm$1.5~km and an albedo of
0.03$\pm$0.01 (assuming $H=13.0$), with a best--fit
$\eta=0.7\pm0.1$. The physical properties derived from the spectrum
agree well with the results obtained from the photometric data in
Section \ref{lbl:thermalmodeling}.

\begin{figure}
\epsscale{0.8}
\plotone{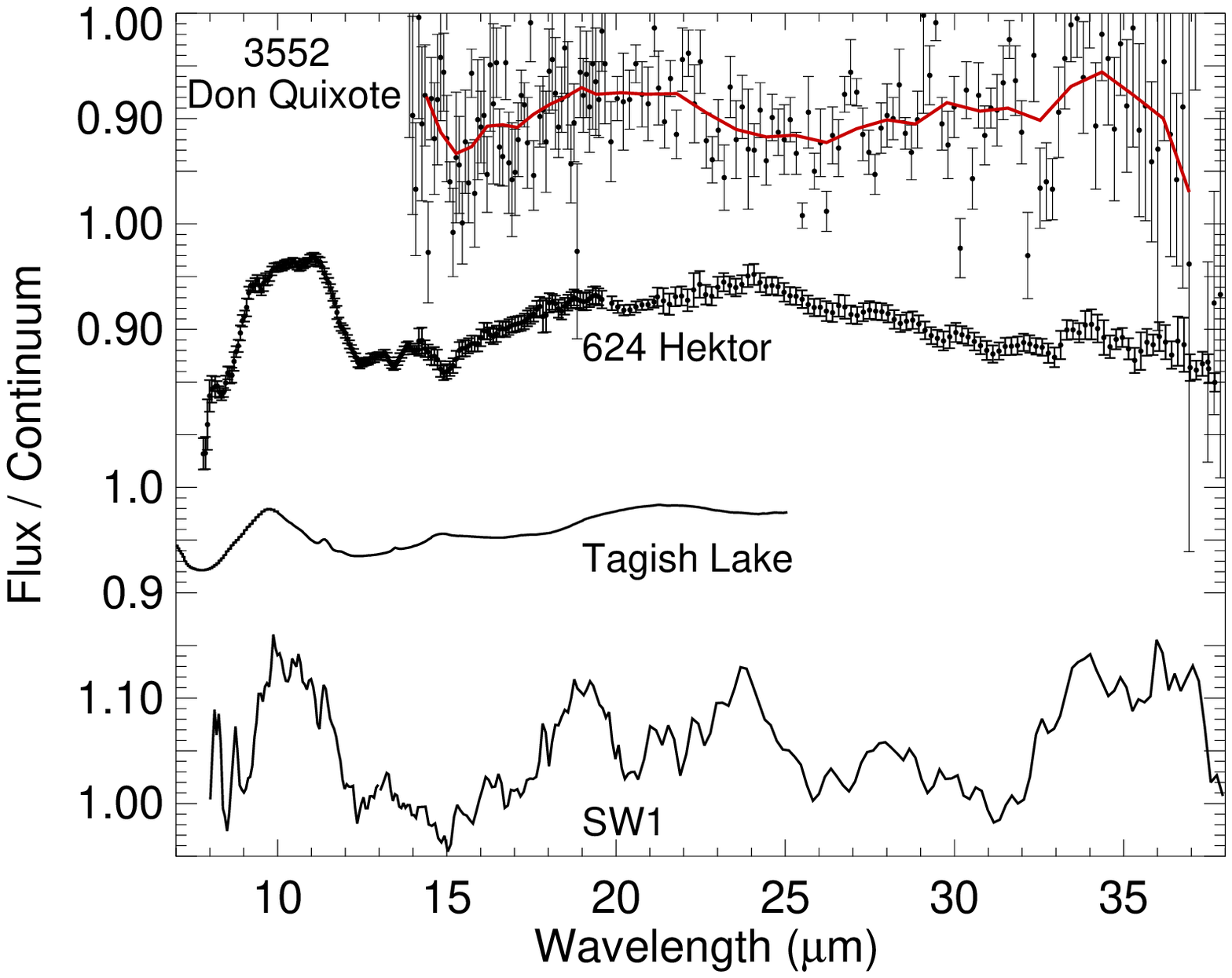}
\caption{Comparison of the Don Quixote IRS spectrum (top) with that of
  D--type asteroid (624) Hektor \citep{Emery2006}, a sample of the
  Tagish Lake meteorite \citep{Hiroi2001}, and short--period comet
  29/P Schwassmann--Wachmann 1 \citep{Stansberry2004}. The continuous
  red line indicates a running average over the Don Quixote spectrum
  and exhibits increased emissivity around 20 and 34~\micron. The
  spectrum of Don Quixote shows significant differences to those of
  the other objects presented here.\label{fig:spectracomparison}}
\end{figure}

Despite the low signal--to--noise ratio of the IRS spectrum of Don
Quixote (Figure \ref{fig:spectracomparison}), clear emissivity peaks
are apparent near 20 and 34~\micron.  Roughly similar emissivity peaks
occur in spectra of comets and primitive asteroids, indicating the
presence of fine--grain silicates.  The Don Quixote spectrum lacks a
peak at 24~\micron\ that is present in many comets and in D--type
Trojan asteroids.  This feature is due to olivine, and we interpret
its absence, on Don Quixote, along with the overall shape and position
of the ${\sim}$20~\micron\ peak, to indicate a more pyroxene--rich
surface than those other objects.


\subsection{Cause and Longevity of the Activity}

The evidence for CO$_2$ band emission implies the existence of
CO$_2$ ice on Don Quixote. The ice is presumably buried under a thick
layer of insulating material \citep{Rickman1990} to explain its
existence in near--Earth space.  Two scenarios can explain the
observed activity: (1) subsurface CO$_2$ sublimates due to seasonal
heating, causing persistent activity, or (2) the observed activity is
temporary, e.g., triggered by a recent impact that exposed icy
sub--surface material, leading to a brief activity outbreak. The
proximity of Don Quixote to the Sun ($r=1.23$~AU) during the IRAC
observations is equally consistent with both of the aforementioned
scenarios. 

Because cometary activity was only unambiguously detected in the IRAC
observations, no direct conclusions regarding the cause and longevity
of the discovered activity can be drawn. However, a comparison with
previous work is informative.  \citet{Kelley2013} found in their
``Survey of the Ensemble Properties of Cometary Nuclei'' (SEPPCON)
that all short--period comets in their sample with $q<1.8$~AU (30
objects) are inactive, presumably because they have already lost all
of their volatiles. The SEPPCON sample targets are known comets that
have shown persistent activity at least in the past. Don Quixote,
showing activity with a perihelion distance of only $q=1.21$~AU,
presents an exception to the findings of \citet{Kelley2013}. Hence, we
speculate that the observed behavior in Don Quixote is most consistent
with a temporary outbreak of activity.

Further observations are necessary to unambiguously constrain the
cause and longevity of the activity. Additional observations during the next
perihelion passage in May 2018 will help to resolve the nature of Don
Quixote's activity.

\subsection{Implications of this Discovery}
\label{lbl:implications}

Because they are spectrally featureless and have no clear meteorite
analog, compositions of D--type asteroids like Don Quixote
\citep{Hartmann1987, Binzel2004} are poorly
constrained. \citep{Vernazza2013} showed that the one meteorite fall
that was originally associated with a D--type spectrum, Tagish Lake
\citep{Brown2000, Hiroi2001}, is not representative for D--type
asteroids. Nevertheless, the low albedo of Don Quixote suggests a
carbonaceous surface material that is generally rich in water and
carbon and at least similar to the Tagish Lake meteorite. Hence, we
use the composition of the Tagish Lake meteorite as an analog for
carbonaceous material. The meteorite has a total water fraction of 3.9
weight percent \citep[wt\%,][]{Baker2002} and a total organic carbon
fraction of 2.6~wt\% \citep{Grady2002}. Assuming a spherical shape for
Don Quixote and a homogeneous composition identical to that of the
Tagish Lake meteorite with a density of 1.5~g~cm$^{-3}$
\citep{Brown2000}, we estimate the total mass of Don Quixote as
${\sim}5\cdot10^{15}$~kg. Based on this mass, the total water content
of Don Quixote would be ${\sim}2\cdot10^{14}$~kg, which equals the
amount of water in the upper--most 1.5~mm of Earth's oceans. The total
organic carbon content of Don Quixote is
${\sim}1.3\cdot10^{14}$~kg. These estimates show that the impact of
such an object could add significant amounts of water and organic
material to Earth's inventory.

Don Quixote has long since been suspected to be of cometary origin as
a result of its comet--like orbit \citep[e.g.,][]{Hahn1985,
  Weissman1989} and albedo \citep{Veeder1989}. Furthermore, dynamical
models clearly suggest a cometary origin \citep{Bottke2002}. The
discovery of activity in this object would not be surprising if the
object had not been lacking any sign of activity in previous
observations. We suppose that Don Quixote's activity evaded discovery
due to either its intermittent nature or, if persistent, the
fact that it is triggered by the sublimation of CO$_2$ ice, the band
emission of which is not observable in the optical.  This assertion is
supported by the low amount of reflected solar light from dust
($Af\varrho\leq4$~cm, which translates into a $V$--band surface
brightness of $\sim$26~mag/square arcsecond), which we derived from the IRAC
3.6~\micron\ flux density. For comparison, the data compiled in
\citet{AHearn1995} show that most known comets with heliocentric
distances comparable to that of Don Quixote have $Af\varrho \sim
100$~cm. If we assume this value to be a rough threshold that triggers
the detection of cometary activity by optical means, the observed dust
production rate, which is linearly related to $Af\varrho$ (see
Equation \ref{eqn:dustproduction}), would have to be at least one
order of magnitude higher.

The existence of CO$_2$ puts constraints on Don Quixote's origin and
evolution: its interior must have formed at very low temperatures
(${\leq}60$~K) to condense CO$_2$ and must have remained cold since
\citep{Yamamoto1985}. The subsurface layers of Don Quixote that
contain CO$_2$--ice are required to have temperatures of 60~K and
below in order to retain the ice. This implies that the diurnal and
seasonal heat waves do not affect the ice layers and are absorbed in
the near--surface insolation layer.

CO$_2$ band emission has been detected in a number of short--period
comets \citep{Ootsubo2012}. Our discovery implies that other NEOs of
cometary origin can retain deposits of CO$_2$ and other volatiles in
the same way as Don Quixote.  Such objects might show temporary
activity and evade discovery by optical means at the same time.
Hence, we suggest expanded monitoring of likely dormant and/or extinct
comets to infrared wavelengths close in time to their perihelion
passage in order to detect CO$_2$ band emission at
4.3~\micron. 

\subsection{Summary}

\begin{itemize}
\item We find evidence for cometary activity in NEO (3552) Don
  Quixote, the third--largest object in near--Earth space, based on
  Spitzer IRAC observations. Extended emission has been detected in
  the 4.5~\micron\ band observations, but only marginally so at
  3.6~\micron. We interpret the lack of a clear detection of a coma at
  3.6~\micron\ as indicating that activity is caused by band emission from
  CO$_2$. The 4.5~\micron\ extended emission shows an anti--sunward
  directed tail with a length of ${\sim}2$\arcmin.
\item From the 3.6~\micron\ band flux--density measurement we
  determine an upper limit on the dust production rate of
  ${\leq}1.9$~kg~s$^{-1}$. Using this estimate and the 4.5~\micron\
  flux density measurement, we constrain a CO$_2$ production rate at
  the time of the Spitzer observations of
    $(1.1\pm0.1)\cdot10^{26}$~molecules~s$^{-1}$.
\item The IRAC observations combined with the additional observations
  from the literature allow for a robust thermal model fit of Don
  Quixote's nucleus, yielding a diameter and albedo of
  18.4$_{-0.4}^{+0.3}$~km and $0.03^{+0.02}_{-0.01}$,
  respectively. Our results confirm that Don Quixote is the
  third--largest known NEO and indicate that it is one of the largest
  known short--period comets.
\item Spectroscopic observations agree with a D--type classification
  of Don Quixote and suggest the presence of fine--grained silicates,
  perhaps pyroxene rich, on its surface.
\item We suspect that Don Quixote's activity has evaded discovery to
  date due to either its possibly intermittent nature or, if
  persistent, the fact that it is triggered by the sublimation of
  CO$_2$ ice, the band emission of which is not observable in the
  optical.
\end{itemize}

\acknowledgments

M. Mommert acknowledges support by the DFG SPP 1385. We thank an
anonymous referee for a number of useful suggestions. This work is
based on observations made with the Spitzer Space Telescope, which is
operated by the Jet Propulsion Laboratory, California Institute of
Technology under a contract with NASA. Support for this work was
provided by NASA through award \#1367413 issued by JPL/Caltech. This
publication makes use of data products from the Wide-field Infrared
Survey Explorer, which is a joint project of the University of
California, Los Angeles, and the Jet Propulsion Laboratory/California
Institute of Technology, funded by the National Aeronautics and Space
Administration.

{\it Facilities:} \facility{Spitzer}.

\clearpage

\end{document}